\documentclass[prl,twocolumn,floatfix,nopacs]{revtex4-1}
\usepackage{graphicx,amsfonts,amssymb,amsmath}
\usepackage[displaymath,mathlines]{lineno}
\usepackage[dvipsnames]{xcolor}
\definecolor{darkblue}{rgb}{0.0, 0.0, 0.75}
\usepackage{color}
\usepackage[colorlinks=true,
            linkcolor=darkblue,
            urlcolor=darkblue,
            citecolor=darkblue]{hyperref}

\newcommand{\subfig}[2]{%
	{\textsf{#1}} \vtop{%
		\vskip0pt
		\hbox{#2}
}}

\newcommand{\brak}[1]{\langle #1\rangle}
\newcommand{\llangle}{\langle\hspace{-0.25em}\langle}
\newcommand{\rrangle}{\rangle\hspace{-0.25em}\rangle}
\newcommand{\minus}{-}
\newcommand{\plus}{+}

\newcommand{\comment}[1]{}

\newcommand*{\bra}[1]{\mathopen{\langle}#1\mathclose{|}}
\newcommand*{\ket}[1]{\mathopen{|}#1\mathclose{\rangle}}
\newcommand{\ketbra}[1]{\ket{#1}\hspace{-0.25em}\bra{#1}}
\newcommand{\ketbrap}[2]{\ket{#1}\hspace{-0.25em}\bra{#2}}
\newcommand{\brav}[1]{\mathopen{\llangle}#1\mathclose{|}}
\newcommand{\ketv}[1]{\mathopen{|}#1\mathclose{\rrangle}}

\newcommand{\braket}[1]{\langle#1\rangle}
\newcommand{\braketp}[2]{\brav{#1} #2\mathclose{\rrangle}}

\begin{document}
	
	\title{Genuine Bistability in Open Quantum Many-Body Systems}
	
	\author{Javad Kazemi}
	\email{javad.kazemi@itp.uni-hannover.de}
	\affiliation{Institut f\"ur Theoretische Physik, Leibniz Universit\"at Hannover, Appelstra{\ss}e 2, 30167 Hannover, Germany}
	\author{Hendrik Weimer}
	\email{hweimer@itp.uni-hannover.de}
	\affiliation{Institut f\"ur Theoretische Physik, Leibniz Universit\"at Hannover, Appelstra{\ss}e 2, 30167 Hannover, Germany}

	\begin{abstract}
		
		We analyze the long-time evolution of open quantum many-body systems
		using a variational approach. For the
		dissipative Ising model, where mean-field theory predicts a wide
		region of bistable behavior, we find genuine bistability only
		at a singular point, confirming the previously suggested
		picture of a first order transition. The situation is dramatically
		different when considering a majority-voter model including
		three-body interactions, where we find bistable behavior in an
		extended region, owing to the breaking of detailed balance in the
		the effective description of the system. In this model, genuine bistability persists even when quantum
		fluctuations are added.
		
	\end{abstract}
	
	\maketitle
	
	
	The metaphorically eternal lifetime of thermodynamically unstable
	diamond gemstones tells us that the minimum free energy is not the
	only thing that matters when considering the long-time properties of a
	physical system. Sufficiently large activation barriers can prevent the
	spontaneous decay of a metastable state even on astronomically large
	timescales. Here we show that similar behavior can also occur for the steady states of open quantum many-body systems, resolving a long-standing controversy about the existence and limits of bistable behavior in these systems.
	
	In the context of open quantum many-body systems, bistable behavior in
	the steady state over an extended parameter range is predicted for
	many different systems by mean-field calculations
	\cite{Lee2011,Lee2012,Marcuzzi2014,LeBoite2013,Jin2013,LeBoite2014,Mertz2016,Parmee2018}. However,
	these findings have so far not held up when using more elaborate
	methods
	\cite{Weimer2015,Weimer2015a,Mendoza-Arenas2016,Maghrebi2016,Kshetrimayum2017,Raghunandan2018,Jin2018,Singh2021}. This
	failure of mean-field theory can be partly understood using the
	variational principle for open systems \cite{Weimer2015}, which allows
	for the formulation of steady state problems in terms of effective
	free energy functionals \cite{Overbeck2017}. In this setting, one can
	see that mean-field solutions for open systems do not correspond to
	the physics above the upper critical dimension, as it is the case for
	equilibrium problems. However, the question whether other open quantum
	systems can support true bistability has remained unsolved.
	
	In this Letter, we present a generic framework for describing
        the long-time evolution of open quantum many-body systems. For
        this, we establish a Gutzwiller approach for open quantum
        systems and employ the variational principle for a mapping
        onto an effective classical problem. In the presence of a
        dynamical symmtetry \cite{Sieberer2013}, fluctuations exhibit
        thermal statistics, allowing us to employ the statistical
        theory of metastability \cite{Langer1969}. For the dissipative
        Ising model exhibiting mean-field bistability
        \cite{Lee2011,Marcuzzi2014}, we demonstrate that the
        thermodynamically unstable solution eventually decays except
        around a singular point, confirming the previously suggested
        picture of a first-order jump in the magnetization
        \cite{Weimer2015,Kshetrimayum2017,Raghunandan2018}. However,
        the situation is fundamentally different when considering a
        majority-voter model known as Toom's model \cite{Toom1980},
        which has recently found applications in topological quantum
        error correction \cite{Herold2017,Kubica2019,Vasmer2021} and
        the realization of time crystals in open systems
        \cite{Zhuang2021}. For this classical spin model, Monte-Carlo
        simulations have reported bistable behavior
        \cite{Bennett1985}, which we also observe within our
        variational approach. We find this behavior being driven by
        the breaking of detailed balance in the non-equlibirium steady
        state, which we analyze based on an effective Langevin
        equation built upon the variational principle. Crucially, we
        also observe that bistability over an extended region persists
        under the inclusion of quantum fluctuations in terms of a
        Hamiltonian driving, constituting the first example of a true
        bistable phase in an open quantum many-body system.

	\emph{Gutzwiller theory for open systems.---}
	The Markovian evolution of quantum states in the form of a density matrix $\rho$ can be described by the Liouvillian superoperator $\partial_t\rho=\mathcal{L}(\rho)$, in terms of the Lindblad master equation 
	\begin{equation}\label{eq:Lindblad}
	\mathcal{L}(\rho)= -i [H,\rho] + \sum\limits_{j}\Big[ {c_j}^\dagger \rho c_j - \frac{1}{2} \big\{ c_j^\dagger c_j, \rho \big\}\Big],
	\end{equation}
	where $H$ is the Hamiltonian of the system and the set of $c_{j}$s
	represents the jump operators \cite{Breuer2002}. This dynamics
	produces non-equilibrium steady states corresponding to
	$\mathcal{L}(\rho_s)= 0$. To approximate these states, we consider a
	Gutzwiller variational ansatz $\rho_v=\prod_i^N \rho_0$, where for
	spin-1/2 particles $\rho_0 \equiv \Big( I + \sum_{\mu
		\in\{x,y,z\}}\alpha_\mu \sigma_\mu \Big)/2$ is expanded in terms of
	Pauli matrices $\sigma_\mu$, while the corresponding coefficients
	$\alpha_\mu$ are a set of variational parameters. These parameters can
	be obtained by minimizing a suitable variational cost function. Here,
	we consider a cost function derived from the vectorized form
	$\braketp{\mathcal{L}^\dagger(\rho)}{\mathcal{L}(\rho)}$ in terms of
	the operator inner product $\text{Tr}\{A^\dagger
	B\}=\braketp{A^\dagger}{B}$ inducing the Hilbert-Schmidt norm
	$||A||_2=\sqrt{\text{Tr}\{A^\dagger A\}}$. However, as the Hilbert-Schmidt norm is biased towards the maximally mixed state \cite{Weimer2015}, we normalize the cost function by the total purity $\text{Tr}\big[\rho^2\big]=\text{Tr}\big[\rho_0^2\big]^N$ to counteract the bias \cite{Vincentini2019}, i.e. $F_v=\braketp{\mathcal{L}^\dagger(\rho)}{\mathcal{L}(\rho)}/\braketp{\rho_0}{\rho_0}^{N}$. 
	
\begin{figure}[t]
		\includegraphics[width=1.0\columnwidth]{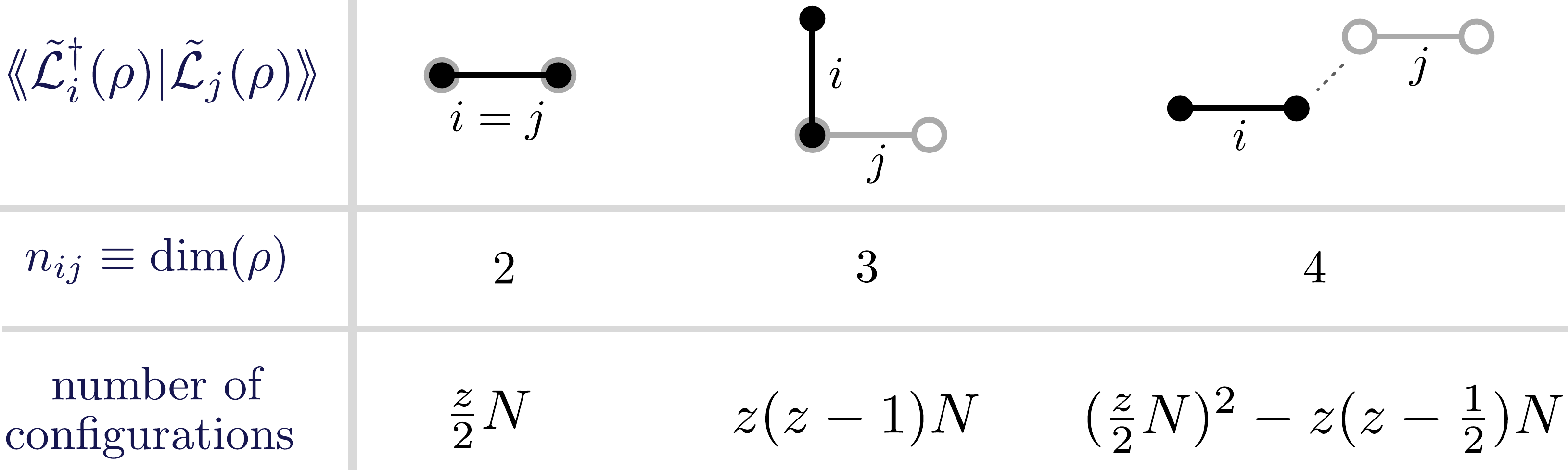}
		\caption{Spatial configurations corresponding to a 2-local Liouvillian appearing on a 2D lattice with the coordination number $z=4$. The black 2-dot lines labelled with $i$ represent $\brav{\tilde{\mathcal{L}}_i^\dagger(\rho)}$ and the gray lines labelled with $j$ represent $\ketv{\tilde{\mathcal{L}}_j(\rho)}$ in the HS norm.}
		\label{fig:Configs}
	\end{figure}

	For a given (uniform) Liouvillian we can express the square of the Hilbert-Schmidt norm $||A||_2^2$ as an expansion of its individual terms, geometrical different configurations as it is shown in fig.~\ref{fig:Configs}, which in turn can be factorized by some $N$- and $N^2$-dependent coefficients. $N$-dependent configurations correspond to those terms in $\braketp{\mathcal{L}^\dagger(\rho)}{\mathcal{L}(\rho)}$ where (fully/partially) mutual lattice sites are acted upon by some elements of $\mathcal{L}^\dagger$ and $\mathcal{L}$ while $N^2$-dependent configurations represent non-overlapping terms. In addition, these coefficients also encode the lattice dimensionality via their dependency on the coordination number $z$. As a further step to have a scale-independent variational norm, which is required in this study, we treat these coefficients on the same footing, rescaling the $N^2$-dependent term to $N$ and then normalize the whole function by $N$. This step can be justified based on the fact that for short-range interacting system the $N$-dependent terms are dominant as they incorporate the overlapping inner products.  
	As a result, the variational norm can be cast into the form
	\begin{equation}\label{eq:v_norm}
	f_v =\sum\limits_{ij} \frac{\braketp{\tilde{\mathcal{L}}_{i}^\dagger(\rho)}{\tilde{\mathcal{L}}_{j}(\rho)}}{\braketp{\rho_0}{\rho_0}^{n_{ij}}},
	\end{equation}
	where $\tilde{\mathcal{L}}_{i}$'s are individual terms in the Liouvillian, similar to what is depicted in fig.~\ref{fig:Configs} and the denominator represents the local purity $\braketp{\rho_0}{\rho_0}$ to the power of $n_{ij}$ being the total dimension of the terms in the numerator.
	
	In cases where the coherent and dissipative couplings are uniform, the variational norm can be evaluated analytically for generic Liouvillians even in the thermodynamic limit,
	analogous to Gutzwiller energies in ground state problems
	\cite{Krauth1992}. This property is in contrast with the more natural
	trace norm $||A||_1=\text{Tr}\{|A|\}$, where additional approximations
	have to be applied to evaluate the variational norm
	\cite{Weimer2015,Weimer2015a}.
	
	\emph{Thermally-activated nucleation processes .---} In cases
        where the variational norm vanishes, the product state
        solution is an exact steady state of the system. Hence, the
        value of the variational norm gives direct access to the scale
        of fluctuations within the system \cite{Overbeck2017}. For the
        first model studied here, these fluctuations obey thermal
        statistics \cite{Sieberer2013}, i.e., they can be captured in
        terms of an effective temperature $T \sim f_v$, considering
        the fact that the variational norm is also an intensive
        quantity.  Importantly, rewriting the steady state problem of
        an open quantum many-body systems in terms of a classical
        statistical mechanics problem allows to employ the vast
        toolbox for treating the latter. This also applies to the
        non-equilibrium relaxation dynamics of metastable states
        \cite{Langer1969}. Here, we first compute all local minima of
        the variational norm. The stability of a local minimim which
        is not the global one (i.e., a metastable state) can then be
        quantified in terms of the relaxation rate by which the
        metastable state relaxes into the global minimum solution.

        To calculate the relaxation rate, we consider the path between
        the metastable minimum and the stable minimum in the
        variational manifold, which is passing through a saddle-point
        having a variational norm of $f_v^{sp}$. Then, in analogy with the statistical theory of the decay of metastable states \cite{Langer1969}, we can express the relaxation rate (per volume) of the metastable solution as 
	\begin{equation}\label{eq:relax_rate}
	I= f_v^{sp} \Big(\frac{f_v^{m}}{2\pi \lambda}\Big)^{1/2} e^{-\tilde{E}_a/f_v^{m}},
	\end{equation}
	where $f_v^m$ is the value of variational norm of the metastable state and $\lambda$ is the curvature of a saddle-point in the activation energy, see below. The equivalent of the activation energy $\tilde{E}_a$ is given here by the variational cost of a critical nucleus of the stable solution (created by random fluctuations) within a system in the metastable state. In the following, we estimate $\tilde{E}_a$ based on classical nucleation theory \cite{Chaikin1995}. By considering square-shaped nucleus with the length $\ell$, we obtain 
	\begin{equation}\label{eq:E_a}
	E_a= -\ell^2 (f_v^m-f_v^s) + 4\ell (f_v^{sp}-f_v^m),
	\end{equation}
	which respectively consists of a volume energy of the nucleus, with $f_v^s$ being the value of variational norm of the stable state, and a surface tension of its domain wall \cite{Chaikin1995}. $\tilde{E}_a$ is then defined as the maximized activation energy with the critical length $\ell^\ast=2(f_v^{sp}-f_v^m)/(f_v^m-f_v^s)$. Concerning the surface tension, we consider a localized sharp kink, in the order of the lattice spacing, separating the stable nucleus from the rest. We have numerically verified that employing a smooth kink using an additional gradient term does not result in any significant change. Finally, $\lambda=2(f_v^m-f_v^s)>0$ is defined as the second derivative of $E_a$ with respect to $\ell$ \cite{Langer1969}.
	
	\begin{figure}[t]
		\begin{center}
			\begin{tabular}{ll}	
				\subfig{a}{\hspace*{-4mm}\includegraphics[width=.46\columnwidth]
					{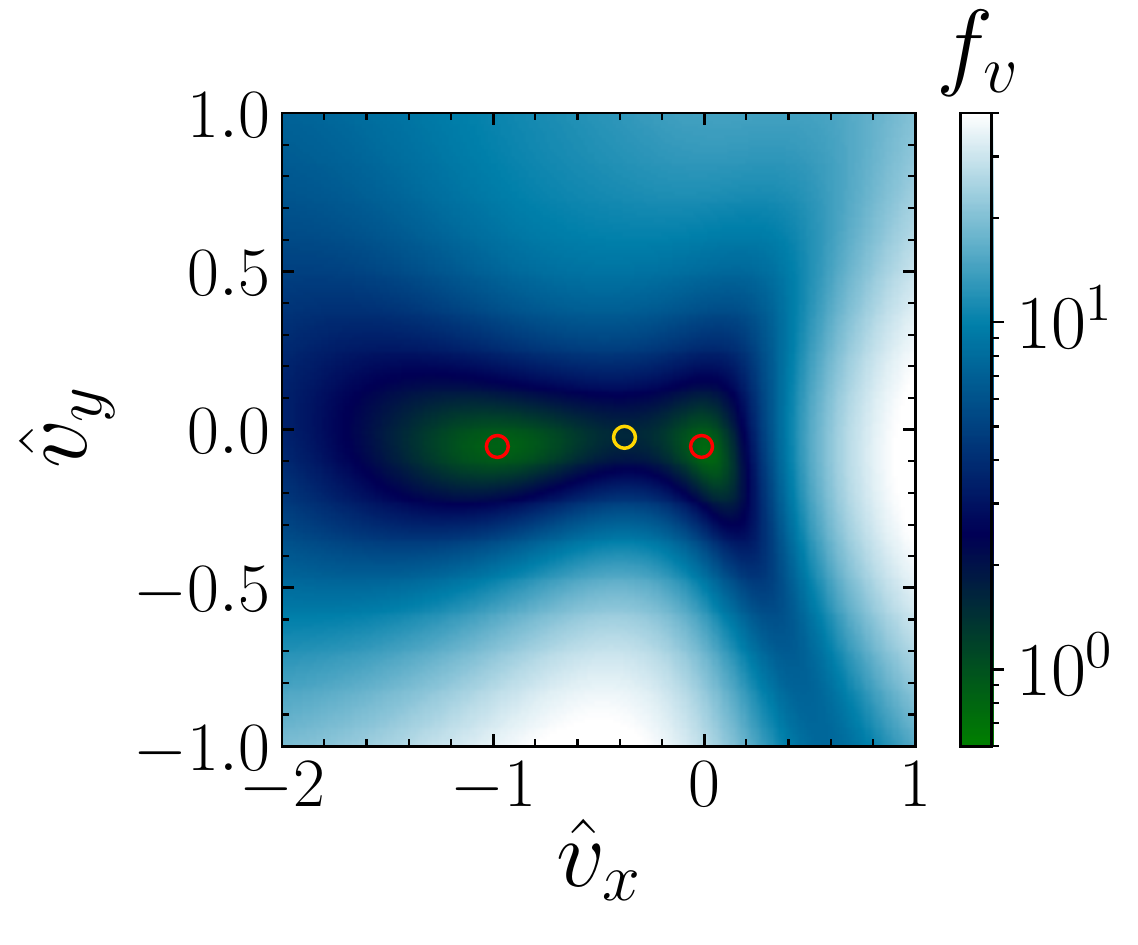}}&
				 \subfig{b}{\hspace*{-4mm}\includegraphics[width=.54\columnwidth]
					{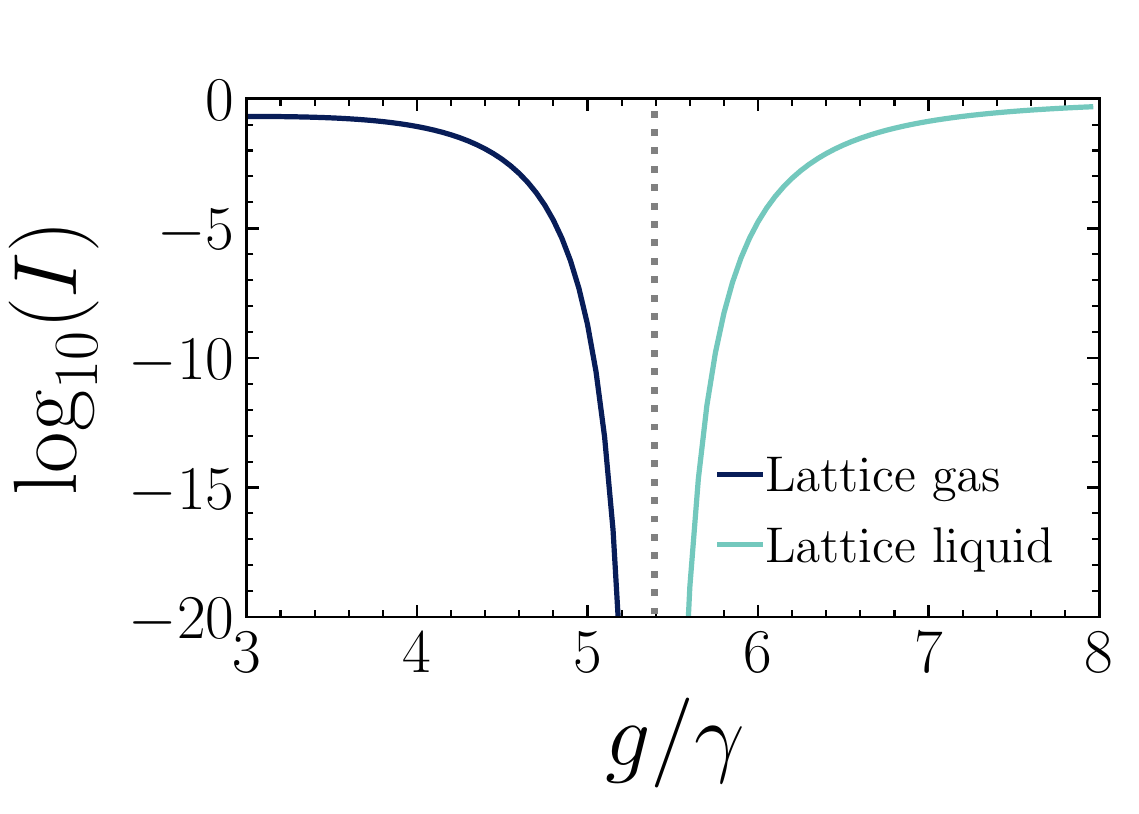}}
			\end{tabular}
		\end{center}
		\caption{(a) Variational norm $f_v$ of the dissipative Ising model at the first-order transition ($J/\gamma=5, g/\gamma=5.40\pm0.01$) in a plane of transformed variational parameters $\hat{v}_x(\alpha_{x,y,z})$ and $\hat{v}_y(\alpha_{x,y,z})$ spanned by the two minima (red circles) and the intermediate saddle-point (yellow circle). The left and right minima correspond to the gas and liquid phase, respectively. (b) Logarithm of the relaxation rate $\log_{10}(I)$ of the metastable solution as a function of $g/\gamma$, showing a steep dip close to the phase transition. }
		\label{fig:Ising}
	\end{figure}
        \emph{Dissipative Ising model.---} As the first concrete model, we turn to the paradigmatic dissipative Ising model, as it is a widely studied model, where mean-field bistability gives way to a first-order transition. Its Hamiltonian part is given by
	\begin{equation}
	H=\frac{g}{2} \sum\limits_{i}\sigma_x^{(i)} + \frac{J}{4} \sum\limits_{\brak{ij}}\sigma_z^{(i)}\sigma_z^{(j)},
	\end{equation}
	where $g$ and $J$ indicate the strength of the transverse
        field and of the Ising interaction, respectively. Dissipation
        is incorporated by adding spin flips in the form of quantum
        jump operators $c_i=\sqrt{\gamma}\sigma_-^{(i)}$ with the rate
        $\gamma$.
	
	Within our variational approach, we can obtain approximative
        steady-state solutions by minimizing the variational norm
        $f_v$ of Eq.~(\ref{eq:v_norm}). Close to the transition, our
        variational results show a double basin structure, see
        Fig.~\ref{fig:Ising}, indicating the presence of two competing
        phases (i.e., liquid and gas), separated by a saddle
        point. We note that our results for the position of the first-order transition are in excellent agreement with other calculations \cite{Weimer2015,Weimer2015a,Kshetrimayum2017,Jin2018,Singh2021}. Crucially, the value of the variational norm at the two
        minima is different except exactly at the first-order transition. To analyze the fate of the metastable state (i.e., the local minimum with the higher variational norm), we compute the relaxation rate defined in eq.~\eqref{eq:relax_rate} to quantitatively assess long-time stability of the metastable solution. As shown in fig.~\ref{fig:Ising}b, the rate is finite far away from the transition and the metastable solution quickly relaxes into the stable one. However, for a narrow width of $g/\gamma$, the relaxation rate of the metastable state drastically decreases, explaining the experimentally observed hysteresis close to the transition \cite{Carr2013}. However, this long-lasting metastability is not a genuine thermodynamic phase, as the relaxation rate is an analytic function over the entire parameter range.
	
	\emph{Toom's majority voting model.---} To answer the question
        whether it is possible to observe genuine bistability in an
        open quantum system, we turn to Toom's majority voting model,
        which supports bistabiliy in a classic non-equilbrium setting
        \cite{Toom1980,Bennett1985,Gacs1988,Grinstein2004}. This model
        has recently found renewed interest because of its relevance
        for topological quantum error correction
        \cite{Herold2017,Kubica2019,Vasmer2021} and time crystals in
        open systems \cite{Zhuang2021}. We explain how genuine
        bistability arises in Toom's model within our variational
        framework and show that this bistability is robust under the
        addition of quantum fluctuations. Toom's model is a set of
        classical rate equations for a system of binary variables
        (i.e., 0 and 1), which can be cast into a fully dissipative
        (i.e., $H=0$) Lindblad master equation, governed by the jump
        operators
        \begin{align}
          c_{j,\mu} = \sqrt{\gamma_\mu} \sigma_-^{j} M_0^{(j,j+E,j+N)}\\
          c_{j,\bar{\mu}} = \sqrt{\bar{\gamma}_\mu} \sigma_+^{j} M_0^{(j,j+E,j+N)}\\
          c_{j,\nu} = \sqrt{\gamma_\nu} \sigma_+^{j} M_1^{(j,j+E,j+N)}\\
          c_{j,\bar{\nu}} = \sqrt{\bar{\gamma}_\nu} \sigma_-^{j} M_1^{(j,j+E,j+N)}
          \label{eq:Toom's_jumps}
          \end{align}
        where $\sigma_{-}=\ketbrap{0}{1}$ and
        $\sigma_{+}=\ketbrap{1}{0}$ are the lowering and raising
        operators, respectively, acting on the site $j$ on a square
        lattice, depending on the state of $j$ and its northern and
        eastern neighbors, expressed in terms of the majority vote
        operators $M_s$. Here, $M_0$ and $M_1$ refer to the
        configurations of the three sites where the majority is in the
        0 and 1 state, respectively, see Tab.~\ref{tab:Toom} for all
        possible configurations. Importantly, rates with index $\mu$
        and $\nu$ refer to lowering and raising operations,
        respectively, while barred and unbarred rates are operations
        against and with the majority, respectively.
        
	
	\begin{table}[b]
		\begin{tabular}{l|cccccccc}\hline \hline
			NCE state & 101 & 111 & 110 & 011 & 010 & 100 & 001 & 000\\\hline
			Operation & $\sigma_\plus$ & $\sigma_\minus$ & $\sigma_\minus$ & $\sigma_\minus$ & $\sigma_\minus$ & $\sigma_\plus$ & $\sigma_\plus$ & $\sigma_\plus$\\\hline
			rate & $\gamma_{\nu}$ & $\bar{\gamma}_{\nu}$ & $\bar{\gamma}_{\nu}$ & $\bar{\gamma}_{\nu}$ & $\gamma_{\mu}$ & $\bar{\gamma}_{\mu}$ & $\bar{\gamma}_{\mu}$ & $\bar{\gamma}_{\mu}$\\\hline\hline
		\end{tabular}
		
		\caption{Ruleset and the corresponding transition rates for Toom's model indicating how the central site is updated according to the North-Center-East (NCE) state.}
		\label{tab:Toom}
	\end{table}
	
	While the deterministic limit of Toom's model,
        i.e. $\bar{\gamma}_\mu = \bar{\gamma}_\nu = 0$ can readily
        eliminate minority islands, Toom has rigorously proved that
        this ability persists even in the presence of updates against the majority rule, provided that the probability of such events is sufficiently low \cite{Toom1980,Lebowitz1990}. This makes Toom's model a fault-tolerant error correcting model for a finite range of the noise \cite{Kubica2019}. Although Toom's original proof is only relevant to the case where the sites are updated synchronously, it has been shown that the argument also holds for simultaneous updates in a master equation formalism \cite{Gray1999}.   
	
	Using a global evolution rate $\gamma$, Toom's model can be characterized in a dimensionless two-parameter space of noise and bias, according to the noise amplitude $T=e^{-\gamma_{\nu}/\gamma}+e^{-\gamma_{\mu}/\gamma}$, analogous to temperature, and with bias $h=(e^{-\gamma_{\nu}/\gamma}-e^{-\gamma_{\mu}/\gamma})/T$, analogous to a symmetry-breaking external magnetic field in the Ising model. In the case of unbiased noise $h=0$, the model behaves like the zero-field Ising model with a continuous transition at a critical temperature. However, even at the presence of biased noise the model undergoes a first-order transition between a bistable phase and a unique ergodic phase. This behavior arises from the chirality of the jump operators in Eq.~\eqref{eq:Toom's_jumps}, as they do not contain the west and south sites \cite{He1990, Munoz2005}. Basically, any chiral updating rule leads to a violation of the detailed balance condition, which is necessary when trying to obtain a stationary state that does not exhibit thermal statistics \cite{Grinstein2004}.

In the presence of thermal statistics, it is relatively
straightforward to express the variatonal norm of
Eq.~\eqref{eq:v_norm} in terms of a Ginzburg-Landau-Wilson framework
and then apply standard techniques for the study of critical phenomena
such as a perturbative renormalization group treatment
\cite{Overbeck2017} to obtain the steady-state phase diagram. However,
we cannot take this route here as the lack of detailed balance can
lead to non-thermal steady states. To overcome this obstacle, we
introduce a Langevin equation to describe the full relaxation dynamics
of the observables $\phi_i = \langle \sigma_z^{(i)}\rangle$. To this end, we
first perform a gradient expansion of the variational norm, i.e.,
$f_v(\{\phi_i\}) = \sum_i f_i(\phi_i, \nabla \phi_i)$, where $\nabla
\phi_i = (\phi_{i+E}-\phi_i, \phi_{i+N}-\phi_i)$ is the lattice
gradient \footnote{See the Supplemental
  Material for the gradient expansion of Toom's model, the detailed derivation of the Langevin equation, and the choice of the initial state.}. The Langevin equation is then given by $\partial_t \phi_i = -\partial f_v/\partial \phi_i + \xi_i$, yielding
	\begin{equation}\label{eq:Langevin}
	\frac{\partial}{\partial t}\phi_i = -\Big[\frac{\partial f_0}{\partial \phi_i} -a- (b-b^\prime) \nabla\phi_i-2b^\prime\phi_i -c \nabla^2\phi_i  \Big] + \xi_i,
	\end{equation}
        where $f_0$ denotes the variational norm for the homogenous
        case without any gradient terms and $\xi_i$ is a white
        Gaussian noise, i.e. $\braket{\xi_i}=0$ and
        $\braket{\xi_i(t)\xi_j(0)}=f_0^m\delta_{i,j}\delta(t)$ with
        $f_0^m$ (the variational norm of the metastable solution) as
        the effective temperature \cite{Note1}. Furthermore, we have truncated the gradient expansion at second order. Our Langevin equation differs from that of conventional kinetic Ising models \cite{Hohenberg1977} due to the appearance of a linear gradient term with coefficient $(b-b^\prime)>0$ which captures the chirality of the jump operators. Here, a sufficiently large $(b-b^\prime)$ ensures the shrinkage of minority islands of either states. Importantly, similar Langevin equations for the Toom's model have already been proposed on a purely phenomenological level \cite{He1990}, while in our variational approach, all coupling constants can be directly calculated from the microscopic model. We also note that our Langevin equation is on equal footing to those that can be derived within the Keldysh formalism \cite{Sieberer2016}, however, performing such calculations for spin systems is often challenging because of the hard-core constraint imposed by spins \cite{Kiselev2000}.
	
	\begin{figure}[t]
		\begin{center}
			\includegraphics[width=.8\columnwidth]
			{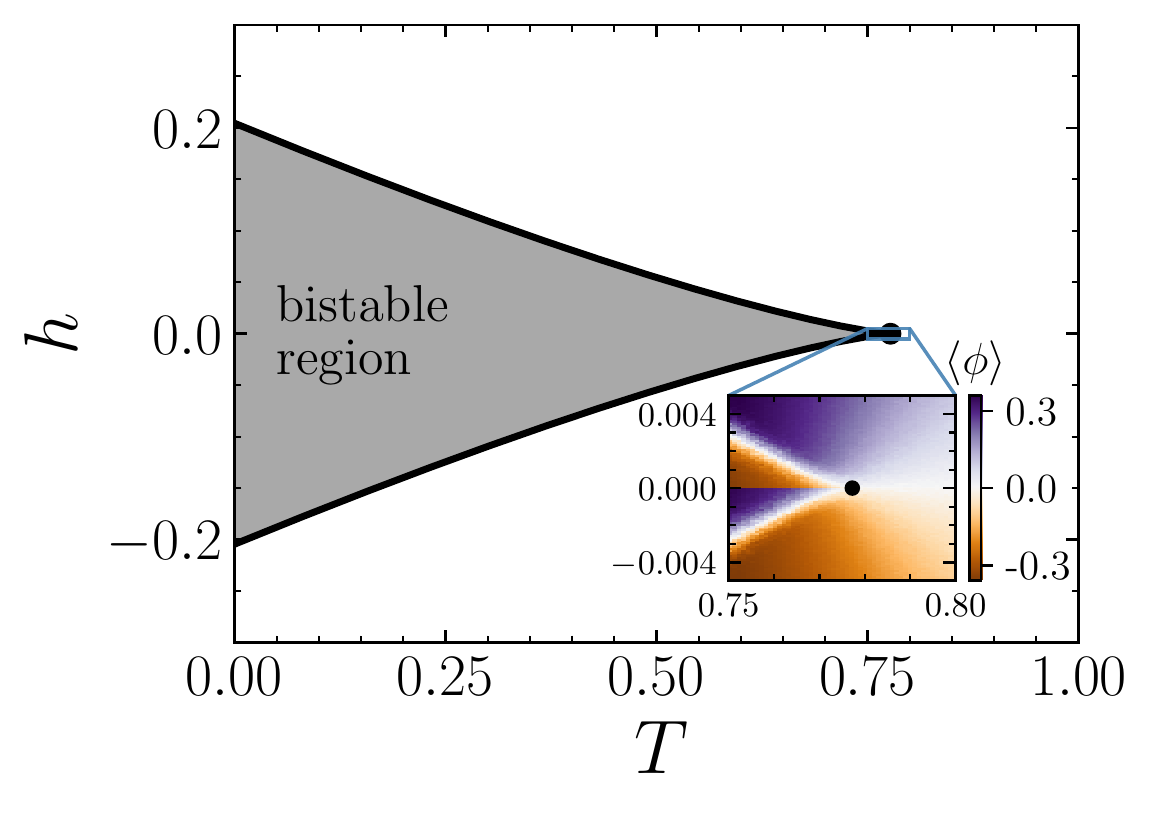}
		\end{center}
		\caption{Phase diagram of The Toom's model shows two phase transition lines separating a bistable phase (gray) from an ergodic phase (white), meeting at a critical point at $T_c=0.777\pm0.001$ (black dot). The inset shows how the orientation of stationary magnetization (starting from a majority of sites polarized against the bias) changes in different zones close to criticality.}
		\label{fig:CToom}	
	\end{figure}

	We are now in the position to calculate the phase diagram of Toom's model in the $T-h$ plane by solving the Langevin equation for a $20\times 20$ lattice and averaging over 100 samples initialized in a configuration polarized against the bias field \cite{Note1}. Fig.~\ref{fig:CToom} demonstrates an extended region of bistability in the absense of the $Z_2$ symmetry of the Ising model that has no counterpart in the corresponding equilibrium system. We note that due to the asynchronicity of master equation in contrast to the Toom's original updating rules, the phase diagram quantitatively differs from that of the synchronous model \cite{Bennett1985}, but the qualitative behavior is identical.  According to the numerical simulation close to criticality, two lines separate the bistable region from the ergodic one ending up to an Ising critical point at $T_c=0.777\pm0.001$ on the $h=0$ line, as shown in the inset of Fig.~\ref{fig:CToom}.

        	\begin{figure*}[t]
		\begin{center}
			\begin{tabular}{llll}	
				\subfig{a}{\includegraphics[width=.65\columnwidth]
					{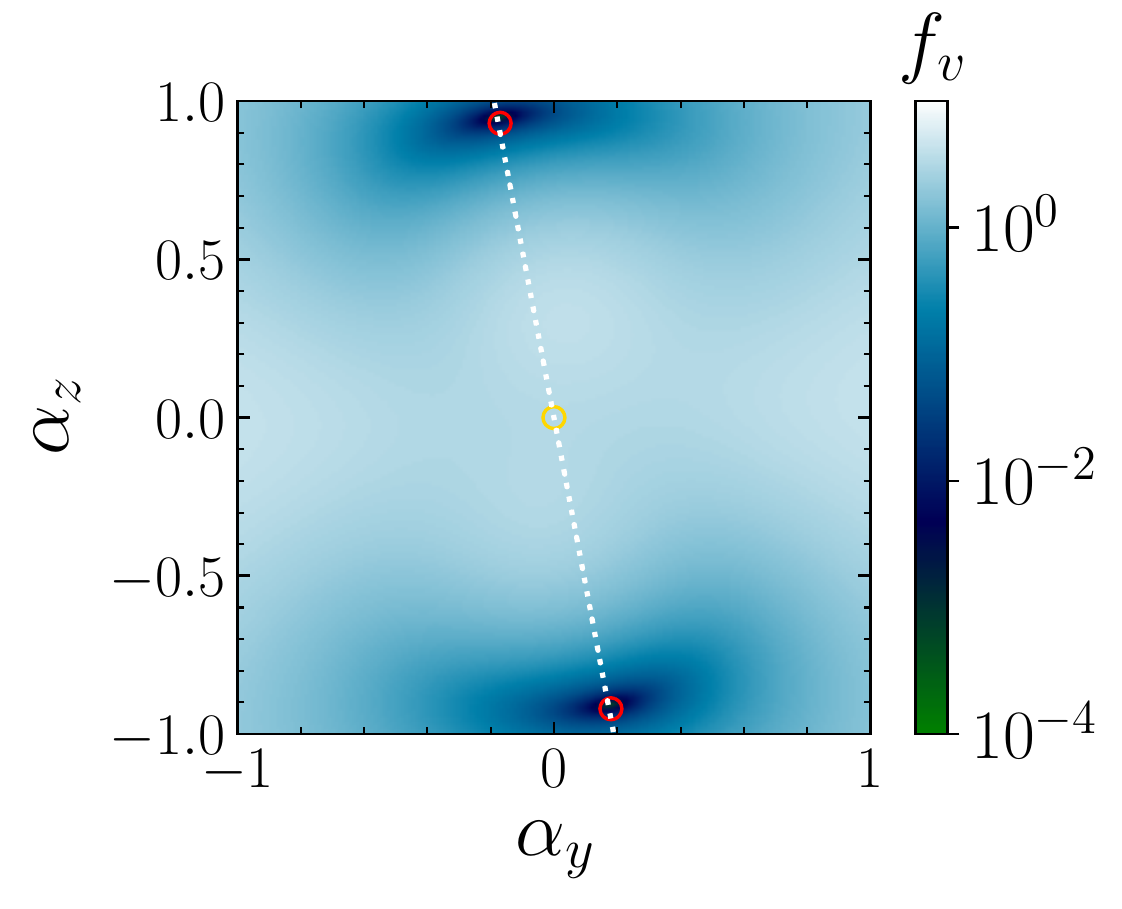}}&
				\subfig{b}{\includegraphics[width=.65\columnwidth]
					{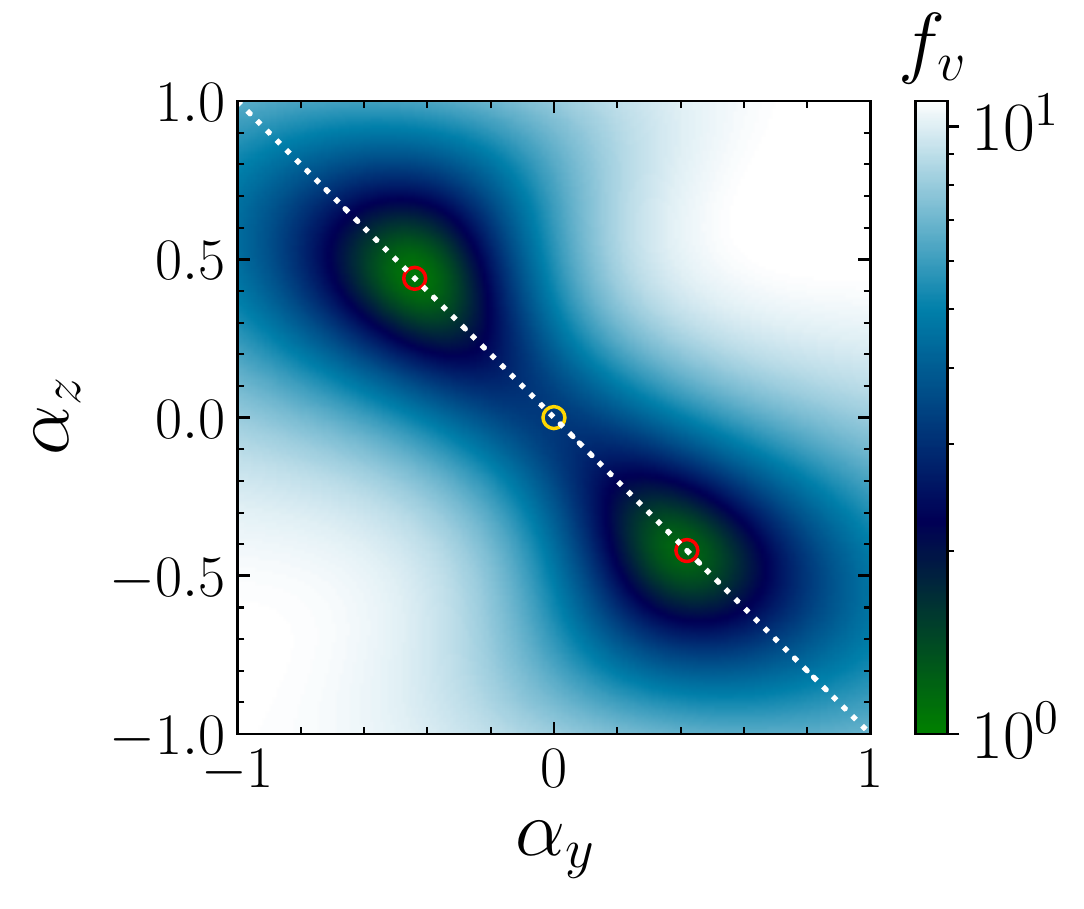}}&
				\subfig{c}{\includegraphics[width=.65\columnwidth]
					{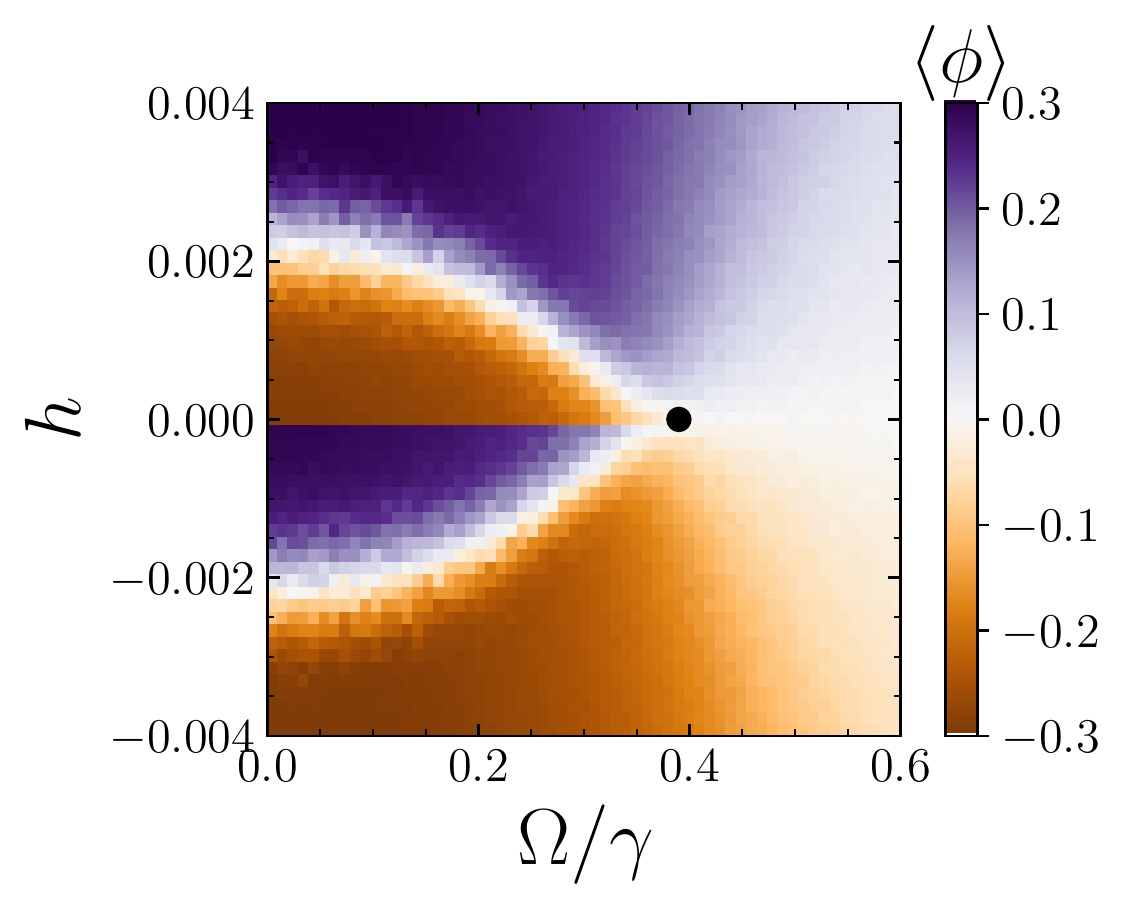}}
			\end{tabular}
		\end{center}
		\caption{Bistability in the variational steady state of Toom's model in the presence of quantum fluctuation. Variational norm (in logarithmic scale) is depicted as a function of $\alpha_z$ and $\alpha_y$ for $H=T=0.15$ deep in (a) the bistable phase with $\Omega/\gamma=0.15$ and (b) the ergodic phase with $\Omega/\gamma=2.0$. The dotted lines show the axis of the effective classical field. (c) Phase diagram in the presence of quantum fluctuations at $T=0.75$ showing gradual shrinkage of the bistable region by increasing $\Omega$ ending up with a critical point (black square) at $\Omega_c/\gamma = 0.39\pm0.02$ similar to the classical case.}
		\label{fig:QToom}
	\end{figure*}

        Having shown that our variational approach is able to
        reproduce the phase diagram of the classical model, we now
        turn to the addition of quantum fluctuations. In the context of Toom's model, this can be done in a natural way in terms of a PXP Hamiltonian of the form
	\begin{equation}
	H=\Omega\sum_{j}\big[P_0^{(j+N)}\sigma_x^{(j)}P_0^{(j+E)}+P_1^{(j+N)}\sigma_x^{(j)}P_1^{(j+E)}\big],
	\end{equation}
        where $P_i = \ketbra{i}$ is a projection operator acting on
        the northern and eastern neighbors. Such PXP terms are of
        great interest in the investigation of strongly interacting
        Rydberg systems \cite{Sun2008,Turner2018,Bluvstein2021}, while the
        realization of the jump operators of
        Eq.~\eqref{eq:Toom's_jumps} is also feasible within these
        systems \cite{Wintermantel2020}. In the language of Toom's
        model, these quantum fluctuations act against the local
        majority and therefore provide a new source of quantum noise
        to the update rules. Additionally, the Hamiltonian conserves
        $\sigma_x$ on all sites, hence the variational Gutzwiller
        ansatz can be parameterized using only $\alpha_y$ and
        $\alpha_z$. We first perform a rotation of the variational
        parameters according to
        $\phi=\cos(\theta)\alpha_z+\sin(\theta)\alpha_y$ with
        $\theta=\arctan(\alpha_y/\alpha_z)$. Importantly, this
        rotation results in $\phi$ containing all the critical
        behavior of the system, while the ortogonal field
        $\phi_{\perp}$ can be approximated by a quadratic term close
        to the variational mimima, see Fig.~\ref{fig:QToom}, i.e., it
        is always gapped. Importantly, while quantum fluctuations lead
        to a renormalization of the coupling constants of the Langevin
        equation for $\phi$, its form remains unchanged.

        To obtain the phase diagram in the presence of quantum
        fluctuations, we first compute the homogeneous variational
        norm $f_v$ in terms of $\alpha_y$ and
        $\alpha_z$. Figs.~\ref{fig:QToom}a and \ref{fig:QToom}b show
        the variational landscape deep in the bistable phase and the
        ergodic phase, respectively. We then compute the phase diagram
        in the $\Omega-h$ plane close to the Ising critical point at
        $T=0.75$, see Fig.~\ref{fig:QToom}c. Crucially, we find
        extended bistability even in the presence of quantum
        fluctuations. Essentially, deep in the bistable region
        (Fig.~\ref{fig:QToom}a), small amounts of quantum fluctuations
        cannot overcome the bistability barrier because the strength
        of the fluctuations indicated by the variational norm $f_v$ is
        very small. However, by increasing $\Omega$
        (Fig.~\ref{fig:QToom}b), both the strength of the fluctuation
        increases substantially and the barrier separating the two
        variational minima is greatly decreased, leading to ergodic
        behavior. Finally, similar to the classical case, the phase
        boundaries separating the bistable from the ergodic phase meet
        in a critical point on the $Z_2$-symmetric line given by
        $h=0$. We have also investigated the phase diagram for
        different Hamiltonian perturbations such as a transverse field
        proportional to $\sigma_x$, where we find qualitatively
        similar behavior. This suggests that Toom's model can serve as
        an error-correcting code in the presence of generic quantum
        fluctuations.
	
	
	To summarize, we have developed a method for assessing the
        long-time evolution of open many-body systems. In the presence
        of a dynamical symmetry giving rise to thermal statistics of
        the steady state, one can calculate the relaxation rate using
        a statistical approach built on classical nucleation
        theory. In the absence of the dynamical symmetry, we find that
        the time evolution of the system can be captured in terms of
        an effective Langevin equation, which allows to map out the
        steady-state phase diagram of systems violating detailed
        balance, where we find the first instance of genuine bistability in an open quantum system when adding quantum fluctuations to Toom's majority voting model. Our approach can be used for many other critical systems that are otherwise inherently difficult to treat, such as quantum contact processes \cite{Carollo2019}, systems exhibiting limit cycles \cite{Owen2018}, or neural networks based on open quantum systems \cite{Rotondo2018}.

	\begin{acknowledgments}
This work was funded by the Volkswagen Foundation, by the Deutsche Forschungsgemeinschaft (DFG, German Research
Foundation) within Project-ID 274200144 -- SFB 1227 (DQ-mat, Project No. A04), SPP 1929 (GiRyd), and under Germany's Excellence Strategy--EXC-2123 QuantumFrontiers--390837967.
	\end{acknowledgments}

%
	
\end{document}


\title{Supplemental Material for ``Genuine Bistability in Open Quantum Many-Body Systems''}
	
	\author{Javad Kazemi}
	\email{javad.kazemi@itp.uni-hannover.de}
	\affiliation{Institut f\"ur Theoretische Physik, Leibniz Universit\"at Hannover, Appelstra{\ss}e 2, 30167 Hannover, Germany}
	\author{Hendrik Weimer}
	\email{hweimer@itp.uni-hannover.de}
	\affiliation{Institut f\"ur Theoretische Physik, Leibniz Universit\"at Hannover, Appelstra{\ss}e 2, 30167 Hannover, Germany}
	\maketitle
	
	\section*{Variational gradient expansion for Toom's model}
	\label{S1}
	
	\begin{figure}[b]
		\includegraphics[width=0.4\columnwidth]{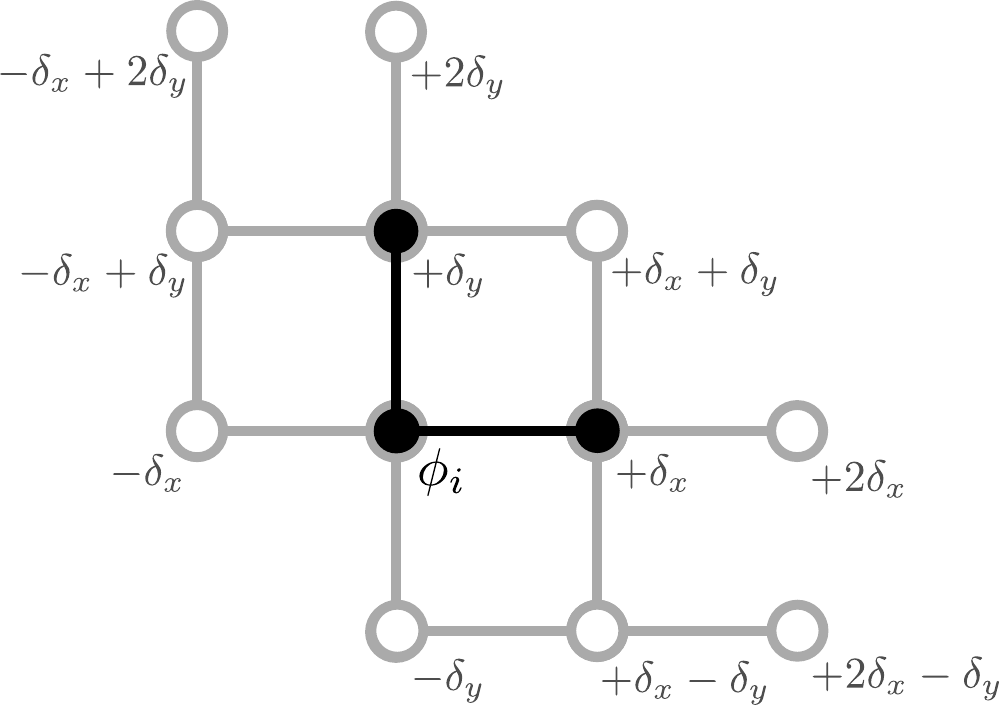}
		\caption{Spatial gradients corresponding to the Toom's model. Here, the black L shape represents $\brav{\tilde{\mathcal{L}}_i^\dagger}$ and the gray L shapes represent $\ketv{\tilde{\mathcal{L}}_j}$ as a function of $\phi_i$ and its spatial gradients $\delta_x$  and $\delta_y$.}
		\label{fig:gradient}
	\end{figure}
	
	We start by formulating a Ginzburg-Landau functional from the variational norm within the Gutzwiller theory for open quantum systems. We consider the functional as a discrete sum over lattice sites including the homogeneous variational norm $f_0$ as well as an inhomogeneous term $f_\nabla$ to incorporates spatial inhomogeneities, i.e.,
	\begin{equation}
	f_v(\{\phi_i\}) =\sum_i f_i(\phi_i,\nabla\phi_i) =f_0+\sum_i f_{\nabla}(\phi_i,\nabla\phi_i),
	\end{equation}
	where $f_{\nabla}$ is constructed by the overlapping terms of the Liouvillian, see Fig.~\ref{fig:gradient}, such that $f_\nabla(\nabla \phi_i = 0)=0$, yielding        
	\begin{equation}
	f_{\nabla}(\phi_i,\nabla \phi_i) = \sum_j\frac{\braketp{\tilde{\mathcal{L}}_{i}^\dagger(\phi_i,\nabla \phi_i)}{\tilde{\mathcal{L}}_{j}(\phi_i,\nabla \phi_i)}-\braketp{\tilde{\mathcal{L}}_{i}^\dagger(\phi_i)}{\tilde{\mathcal{L}}_{j}(\phi_i)}}{\braketp{\rho_0(\phi_i)}{\rho_0(\phi_i)}^{n_{ij}}},
	\end{equation}
	with $j$ running over both $i$ and the nearest neighbors that
        are coupled via the action of the Liouvillian. We can  express $f_\nabla$ in terms of the localized field $\phi_i$ and the spatial inhomogeneities around it, i.e. $\nabla\phi_i=\delta_x\hat{x}+\delta_y\hat{y}$, with
	\begin{equation}
	\begin{split}
	\delta_x=&\big(\phi_{i+E}-\phi_{i}\big)=-\big(\phi_{i+W}-\phi_{i}\big)\\
	\delta_y=&\big(\phi_{i+N}-\phi_{i}\big)=-\big(\phi_{i+S}-\phi_{i}\big),
	\end{split}
	\end{equation}
        where the capital letters denote the cardinal direction. For Toom's model, performing a Taylor expansion of $f_\nabla$ around $\phi_i=0$ up to the second order yields
        \begin{equation}
		f_\nabla = \frac{a}{2}\big[\delta_x+\delta_y\big]+b\delta_x\delta_y+b^\prime\phi_i\big[\delta_x+\delta_y\big]+c\big[\delta_x^2+\delta_y^2\big],
		\end{equation}
        where $a$, $b$, $b^\prime$, and $c$ are the expansion coefficients of the variational norm. Using lattice gradients, this can be expressed as
       	\begin{equation}
       f_\nabla = \left(\frac{a}{2}+b^\prime\phi_i\right)\nabla\phi_i+b\left(\nabla\phi_i\cdot\hat{x}\right)\left(\nabla\phi_i\cdot\hat{y}\right)+c\left(\nabla\phi_i\right)^2.
       \end{equation}

	\section*{Derivation of the variational Langevin equation}

        To obtain the time evolution of a classical field $\phi(t)\equiv \phi_t$, we consider a generic density matrix $\rho(t)$ as a function of a small perturbation $\epsilon$
 	\begin{equation}
	  \rho(t) =\frac{1}{2}\big[\rho(\phi_t+\epsilon)+\rho(\phi_t-\epsilon)\big].
          \label{eq:rhoeps}
	\end{equation}
        In the following, we consider the decay of the variational trial
        state $\rho$ under the Lindbladian $\mathcal{L}$, i.e.,
        \begin{equation}
          \rho(t+\tau) = \exp(\mathcal{L}\tau)\rho(t) \approx \mathcal{N} \exp(-f_v\tau)\rho(t),
        \end{equation}
        where $\mathcal{N}$ is a normalization factor. Inserting Eq.~(\ref{eq:rhoeps}) yields	
	\begin{equation}
	\rho(t+\tau)=\frac{e^{-f_v(\phi_t+\epsilon)\tau}\rho(\phi_t+\epsilon)+e^{-f_v(\phi_t-\epsilon)\tau}\rho(\phi_t-\epsilon)}{e^{-f_v(\phi_t+\epsilon)\tau}+e^{-f_v(\phi_t-\epsilon)\tau}},
	\end{equation}
        or equivalently, for the field $\phi$
	\begin{equation}
	\phi_{t+\tau}=\frac{e^{-f_v(\phi_t+\epsilon)\tau}(\phi_t+\epsilon)+e^{-f_v(\phi_t-\epsilon)\tau}(\phi_t-\epsilon)}{e^{-f_v(\phi_t+\epsilon)\tau}+e^{-f_v(\phi_t-\epsilon)\tau}}
	\end{equation}
	Performing a Taylor expansion around $\phi$ up to the second order, we arive in the limit $\tau \to 0$ at
	\begin{equation}
	  \frac{\partial}{\partial t}\phi =-\epsilon^2 \frac{\partial f_v}{\partial \phi}.
	\end{equation}
        Here, $\epsilon$ can be interpreted to define a mesoscopic
        timescale. We can express the evolution of $\phi$ in the units
        of the mesoscopic timescale by performing the rescaling
        $\epsilon^2 t \to t$. Additionally, it is straightforward to
        generalize this expression to many-body systems containing
        multiple fields $\phi_i$.

        Importantly, the rescaled equation of motion for $\phi_i$ is
        not yet self-consistent as it lacks fluctuations. For
        self-consistency of the solution, the strength of fluctuations
        needs to correspond to the variational norm of the homogeneous
        solution \cite{Overbeck2017}. This can be achieved by adding a
        stochastic term $\xi$ to the equation of motion, i.e.,
	\begin{equation}
	\frac{\partial}{\partial t}\phi_i =-\frac{\partial f_v}{\partial \phi}|_{\phi=\phi_i}+\xi_i,
	\end{equation}
        where $\xi_i$ is white Gaussian noise having zero mean and a
        correlation function given by
        $\braket{\xi_i(t)\xi_j(t')}=f_h(\phi_i)\delta_{i,j}\delta(t-t')$,
        with $f_h(\phi_i)$ denoting the variational norm of the
        homogeneous solution according to $\phi_i$. Close to
        criticality, $f_h(\phi_i)$ can be replaced by the variational
        minimum $f_0$. Then, the Langevin equation for Toom's model is given by
	\begin{equation}
	\frac{\partial}{\partial t}\phi_i = -\Big[\frac{\partial f_0}{\partial \phi_i} -a- (b-b^\prime) \nabla\phi_i-2b^\prime\phi_i -c \nabla^2\phi_i  \Big] + \xi_i,
	\end{equation}
	 where the gradient terms indicate discrete lattice differences according to
	 \begin{eqnarray}
	 \begin{split}
	 \nabla\phi_i&=\phi_{i+N}+\phi_{i+E}-2\phi_{i},\\
	 \nabla^2\phi_i&=\phi_{i+N}+\phi_{i+S}+\phi_{i+W}+\phi_{i+E}-4\phi_i.
	 \end{split}
	 \end{eqnarray}
        
	\section*{Intial state for variational simulations}
	\label{S2}

\begin{figure}[t]
		\begin{center}	
				\includegraphics{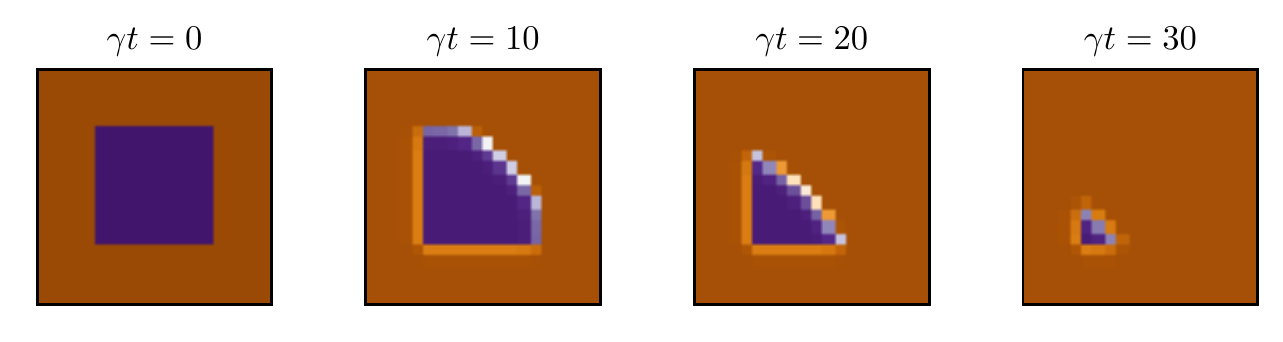}
		\end{center}
		\caption{Time evolution of a $10\times10$ minority island of the stable state encircled by the less-stable state on $20\times20$ square lattice with $T=0.75$, $h=0.001$ and zero white noise.}
		\label{fig:triangle}	
	\end{figure}

        The bistability within Toom's model can be traced back to its
        ability to eliminate minority islands of arbitrary size. This
        elimination process is also independent of the island's
        shape. In Fig.~\ref{fig:triangle}, we illustrate this by
        solving the equation of motion for a lattice of $20\times 20$
        sites. As the initial state in the simulation, we consider a
        square-shape minority island with a length of $10$ whose state
        is favored by the bias field $h$ and the bulk being in the
        unfavored state. We observe that in the case of zero white
        noise, the square-shape island gradually turns into a triangle
        and then the aforementioned shrinkage takes place. While the
        white noise is very small only deep in the bistable region,
        the principal argument also holds close to criticality where
        the noise terms become important. Since all initial shapes
        are reduced to triangles on timescales much faster than
        the overall relaxation, we consider initial states that
        have a triangular minority island polarized along the bias
        field, while the majority of the initial state is polarized
        against the bias. Specifically, we choose a triangle with a side length of $10$ sites in a $20\times20$ lattice with periodic boundary conditions. We then solve the stochastic Langevin equation and average over 100 realizations.
